\documentclass{article}

\usepackage{PRIMEarxiv}

\usepackage[utf8]{inputenc} 
\usepackage[T1]{fontenc}    
\usepackage{hyperref}       
\usepackage{url}            
\usepackage{booktabs}       
\usepackage{amsfonts}       
\usepackage{nicefrac}       
\usepackage{microtype}      
\usepackage{lipsum}
\usepackage{gensymb} 
\usepackage{fancyhdr}       
\usepackage{graphicx}       
\usepackage{pifont}
\graphicspath{{media/}}     

\usepackage{siunitx}
\usepackage{dsfont}
\usepackage{subcaption}
\usepackage{lineno}
\usepackage{bm}
\usepackage{amsmath}
\usepackage{natbib}
\usepackage{authblk}
\usepackage{titlesec}
\usepackage{amssymb}
\usepackage{comment}
\usepackage{soul}

\title{Forecast error diagnostics in neural weather models}

\author[1]{\textbf{Uro\v s Perkan}}
\author[2,1]{\textbf{\v{Z}iga Zaplotnik}}
\author[1]{\textbf{Gregor Skok}}

\affil[1]{University of Ljubljana, Faculty of Mathematics and Physics, \protect\\Jadranska 19, 1000 Ljubljana, Slovenia\vspace{2mm}}
\affil[2]{European Centre for Medium-Range Weather Forecasts, \protect\\Robert-Schuman-Platz 3, 53175 Bonn, Germany}

\date{} 

\usepackage{tikz}

\usepackage{atbegshi} 

\newcommand{\firstpagefooter}{%
  \begin{tikzpicture}[remember picture, overlay]
    \node at ([yshift=1.5cm]current page.south) {
      \begin{minipage}{\textwidth}
        \centering
        \rule{\textwidth}{0.2pt}\\
        \scriptsize{\textit{Corresponding authors:} Uro\v s Perkan (\texttt{uros.perkan@fmf.uni-lj.si})}
      \end{minipage}
    };
  \end{tikzpicture}%
}

\AtBeginShipoutFirst{\firstpagefooter}

\begin{document}
\maketitle

\begin{abstract}
Deep-learning (DL) weather prediction models offer some notable advantages over traditional physics-based models, including auto-differentiability and low computational cost, enabling detailed diagnostics of forecast errors. Using our convolutional encoder-decoder model, ConvCastNet, we systematically relax selected subdomains of the forecast fields towards "true" weather states (ERA5 reanalyses) and monitor the forecast skill gain in other regions. Our results show that a medium-range mid-latitude forecast improves substantially when the stratosphere and boundary layer are relaxed, while relaxation of the tropical atmosphere has a negligible effect. This underscores the need for a more accurate representation of the stratosphere and the planetary boundary layer to improve medium-range weather predictability. Additionally, we investigate the relationship between the forecast error sensitivity to initial conditions and relaxation experiments. By utilising auto-differentiability, we identify overlapping regions of large error sensitivity and strong forecast skill improvement from relaxation. Average mid-latitude error sensitivity to initial conditions shows negligible influence from the tropics, corroborating the results of the tropical relaxation experiments. The error sensitivity shows a physically consistent influence of upstream weather dynamics and sea surface temperatures on forecast accuracy. The latter also highlights the importance of accurately representing the atmosphere–ocean coupling in numerical weather prediction models. This combined approach could provide valuable heuristics for diagnosing neural model errors and guiding targeted model improvements.
\end{abstract}




\section{Introduction}\label{sec1}

Extensive research has established data-driven deep-learning (DL) weather prediction models as fast, energy-efficient, and skilful tools for short- and medium-range weather prediction. Some of the most prominent architectures include convolutional neural networks (CNN) \citep{Weyn2019, Weyn2020, Weyn2021}, graph neural networks \citep{Keisler2022GNN, GraphCast, AIFS}, transformers \citep{ClimaX, Pangu-Weather, FuXi, Aurora}, Fourier neural operators \citep{SFNO} and hybrids \citep{FourCastNet, GenCast}. Regional DL models have also outperformed physics-based models according to some standard metrics \citep{MetNet3}. Even though data-driven DL models have been shown to be superior to physics-based models in some aspects, they are distinct from their physics-based counterparts in their lower prior physical knowledge and weaker inductive biases, making them susceptible to predicting non-realistic physical balances, larger smoothing of predicted fields and physically inconsistent energy spectra \citep{Bonavita2024}. In contrast to physics-based models, where underlying physical laws and numerical discretisation schemes offer interpretability of the simulated processes and potential sources of errors, data-driven DL models lack explicit mechanisms for explaining their predictions. Therefore, understanding and attributing forecast errors is essential for enhancing the performance of DL-based weather forecasting systems.

Forecast errors can arise from poor initial conditions and grow due to chaotic weather dynamics. Error growth is further accelerated due to model limitations, which expedite the divergence of the actual and forecasted trajectories. In classical models, this includes errors associated with numerical and physical approximations \citep{QuietRevolution}, which become most evident in intense cyclogenesis, tropical-to-extratropical cyclone transitions, mesoscale convective systems in the extratropics, organised deep convection in the tropics, etc. DL model errors, in contrast, arise due to limitations of the specific model architecture, loss function and minimisation algorithm \citep[e.g.][]{DoublePenaltyLoss}, as well as by inheriting training data errors and biases. In physics-based models, error diagnostics can be performed using several methods. Some of the typically used include plotting the forecast error against analysis for different lead times, applying ensemble sensitivity methods \citep[e.g.][]{EnsembleSensitivity} or using the relaxation (nudging) technique \citep[e.g.][]{Relaxation2014, DiagnosticMethods}. 

An essential difference between DL and physics-based models is that modern DL libraries allow DL models to be programmed in an auto-differentiable manner. This means we can easily compute the gradient of the model's output with respect to the input fields. Such gradients, often referred to as saliency maps \citep{SaliencyMaps}, are a tool from explainable artificial intelligence (AI) that can be adapted to identify the regions of the input fields to which the forecast error—measured over a selected set of output variables and grid points—is most sensitive. This method closely resembles the adjoint sensitivity technique \citep[e.g.][]{WhatIsanAdjointModel}, the difference being that the DL model is fully non-linear. In recent studies, gradient-based sensitivity analyses have been used to compute the sensitivity of mid-latitude and equatorial prediction to the input geopotential \citep{SaliencyMapsCNN}, the sensitivity of the total precipitation (TP) anomaly to the sea surface temperatures (SST) in Pakistani floods \citep{PakistanFloods}, saliency analysis of atmospheric rivers \citep{AtmosphericRiver} or finding the optimal initial conditions to minimise the North American heatwave forecast error \citep{PredictabilityLimit}. Gradient-based sensitivity analysis could be a helpful tool for identifying potential problems, such as model predictions being influenced by processes which, based on our physical understanding, should not play a role (like the overreaching region of influence in 850 hPa temperature prediction found by \citep{SaliencyMapsCNN}. It also holds potential for advancing the discovery of new physics.

The origins of forecast errors in DL weather prediction models remain insufficiently understood. Advancing this understanding could accelerate model development. In this study, we investigate the physical consistency of error sensitivity maps produced by our convolutional encoder-decoder model, ConvCastNet, and introduce a relaxation method that locally nudges predicted fields towards the reanalysed ("true") weather state. We analyse how the nudging affects subsequent forecasts and demonstrate how the relaxation technique is inherently linked to the model's sensitivity to input fields. Combining these two approaches could provide valuable insights into the sources of forecast errors in DL-based weather prediction. 

The structure of the paper is as follows. Section \ref{sec: ConvCastNet} describes the ConvCastNet model. In Section \ref{sec: Methodology}, we describe the methodology for gradient-based sensitivity calculation and relaxation experiments. Results of those experiments are described in Section \ref{sec: Results}, while Conclusions and Outlook are given in Section \ref{sec: Conclusions}. 

\section{ConvCastNet}\label{sec: ConvCastNet}

\subsection{Model architecture}\label{sec: Model architecture}

\begin{figure*}
\centerline{\includegraphics[width=1. \textwidth]{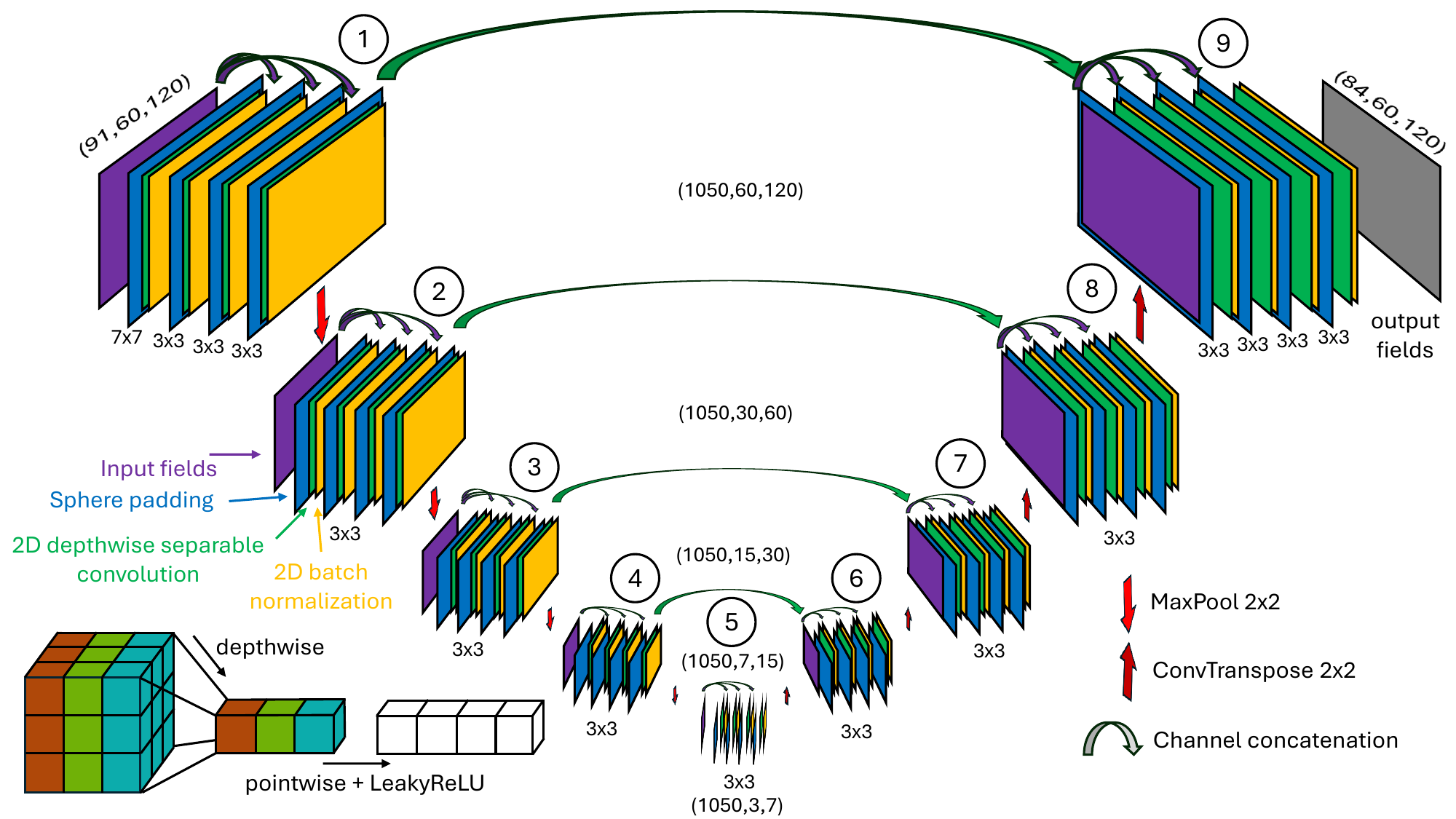}}
\caption{Schematics of ConvCastNet architecture. Coloured layers, positioned in the 9 model blocks, represent input and latent tensors. Specifically, violet colour layers represent input tensors of a given block, dark blue layers show spherically padded tensors, green layers are produced by DS convolutions with Leaky ReLU activation, and orange layers are batch-normalised outputs of green layers. DS convolutions are illustrated in the bottom left corner. The left cube represents a part of the input tensor covered by the DS convolutional kernel of size $(3,3,3)$; the middle column shows features after the depthwise convolution, where a single filter is applied independently to each input channel, and the right column shows features after the pointwise ($1\times 1$) convolution and activation. Max pool and transposed convolution operations are shown using down and up arrows, respectively. Skip connections are plotted as curved arrows and show which tensor is concatenated to which in the channel dimension. Intermediate tensor dimensions are printed above skip connections and below the bottom-most layer. The output fields are plotted in grey. The input tensors to the ConvCastNet have shape (91,60,120), while the output tensors have shape (84,60,120), the input having more channels due to additional input fields (see Table~\ref{tab: data}). \label{fig: ConcCastNet}}
\end{figure*}

The experiments described in this paper are performed using the convolutional encoder-decoder neural network, ConvCastNet (Figure~\ref{fig: ConcCastNet}), based on the U-Net architecture \citep{UNet}. It consists of 9 neural network blocks, each comprising 4 sub-blocks, where depthwise separable (DS) convolutional layers \citep{DepthwiseSeparableConv} are applied to spherically padded input fields. In these layers, the convolution is split into two steps: a depthwise convolution, where a spatial filter is applied independently to each input channel, and a pointwise convolution, where a $1\times1$ convolution combines the output across channels. Each convolutional layer is followed by a leaky rectified linear unit (Leaky ReLU) activation and 2D batch normalisation. Blocks are separated by max pooling or transposed convolution operators, and the output fields are obtained from the last block using the 2D pointwise convolution. DS convolutions are used due to their ability to learn both spatial and channel-wise connections while significantly reducing the number of trainable parameters, compared to classical convolutional layers. This reduces the model’s complexity, enhances computational efficiency, and improves generalisation beyond the training data. Further details of the ConvCastNet deep learning model are provided in \ref{App: TechnicalDetails}. 

\subsection{Training data}

The data for neural-network training, validation and testing is obtained from 53 years (1970-2022) of ERA5 reanalysis \citep{ERA5}, using the 00 and 12 UTC time instances. It is interpolated to a 3-degree lat-lon grid between latitudes $\varphi = \pm88.5\degree$ using a bicubic spline interpolator \citep{BicubicSpline}. Data variables and properties are listed in Table~\ref{tab: data}. The input to the model consists of 15 variables, 8 of which are prognostic. The sea-ice cover (sic), snow and volumetric soil water (vsm) are persistent throughout the forecast rollout. Their inclusion as static variables likely improves the medium-range weather forecast; however, it makes the model less suitable for long-range forecasting \citep{Cohen2019}. Latitudes, land-sea mask, and elevation above sea level are also used as static variables, while top-of-atmosphere incident solar radiation (TOASR) changes spatially and temporally, consistent with the date and time of the forecast lead time. The pressure level variables are predicted at 13 pressure levels. 

\begin{table*}[t]%
\centering
\caption{Meteorological data used for ConvCastNet training. Note that 50, 100, ..., 1000 denotes 13 pressure levels: 50, 100, 150, 200, 250, 300, 400, 500, 600, 700, 850, 925, 1000. \label{tab: data}}%
\resizebox{\textwidth}{!}{%
\begin{tabular*}{\textwidth}{@{\extracolsep\fill}llllll@{\extracolsep\fill}}%
\toprule
\textbf{Variable name} & \textbf{Annotation} & \textbf{Single level} & \textbf{Pressure levels} & \textbf{Input} & \textbf{Output} \\
\midrule
geopotential & $\phi$ & & 50, 100, \dots, 1000 & \ding{51} & \ding{51}\\
specific humidity & $q$ & & 50, 100, \dots, 1000 & \ding{51} & \ding{51}\\
temperature & $T$ & surface, 2 $\mathrm{m}$ & 50, 100, \dots, 1000 & \ding{51} & \ding{51}\\
zonal wind & $u$ & 10 $\mathrm{m}$ & 50, 100, \dots, 1000 & \ding{51} & \ding{51}\\
meridional wind & $v$ & 10 $\mathrm{m}$ & 50, 100, \dots, 1000 & \ding{51} & \ding{51}\\
vertical velocity & $\omega$ & & 50, 100, \dots, 1000 & \ding{51} & \ding{51}\\
mean sea level pressure & $\mathrm{mslp}$ & & & \ding{51} & \ding{51}\\
total precipitation & $\mathrm{prec}$ & & & \ding{51} & \ding{51}\\
sea-ice cover & $\mathrm{sic}$ & & & \ding{51} & \\
snow depth & $\mathrm{snow}$ & & & \ding{51} & \\
volumetric soil water & $\mathrm{vsm}$ & & & \ding{51} & \\
TOA incident solar radiation & $\mathrm{toasr}$ & & & \ding{51} & \\
latitudes & $\mathrm{\varphi}$ & & & \ding{51} & \\
land sea mask & $\mathrm{lsm}$ & & & \ding{51} & \\
elevation & $z$ & & & \ding{51} & \\
\bottomrule
\end{tabular*}
}
\end{table*}

Data is normalised as
\begin{equation}\label{eq: standardisation}
\mathcal{S}(X) = \frac{X - \bar{X^c}}{\mathrm{std}(X^c) + \varepsilon} \, ,  
\end{equation}
where $\bar{X^c}$ denotes the climatological mean, $\mathrm{std}(X^c)$ is the climatological standard deviation, and $\varepsilon=10^{-7}$ is a small constant added for numerical stability. The climatological mean and standard deviation are computed over the period 1950-2014. With the exception of sea-ice cover, snow and volumetric soil moisture, where the mean and standard deviation are computed globally by aggregating over latitude and longitude, standardisation is performed separately for each latitude, longitude and pressure level.

Predictions are rolled out autoregressively as $\hat{y}_{t+1} = \mathcal{M}(\hat{y}_{t})$, where $\hat{y}_{t}$ is the standardised prediction at time step $t$ and $\hat{y}_{0} = y_{0}$. The model is trained to predict time steps with $\Delta t = 12$ hours.

The data is separated into training, validation and testing subsets, with the latter covering the 2020-2022 time span. The number of autoregressive steps $n$, the number of epochs and the time interval for training and validation subsets are printed in Table~\ref{tab: training}. At each stage, we save the model with the highest anomaly correlation coefficient (ACC) on the validation dataset and use it to initialise the weights for the next stage.

\subsection{Forecast skill}\label{subsec: Forecast skill}

\begin{figure*}
\centerline{\includegraphics[width=1.13\textwidth]{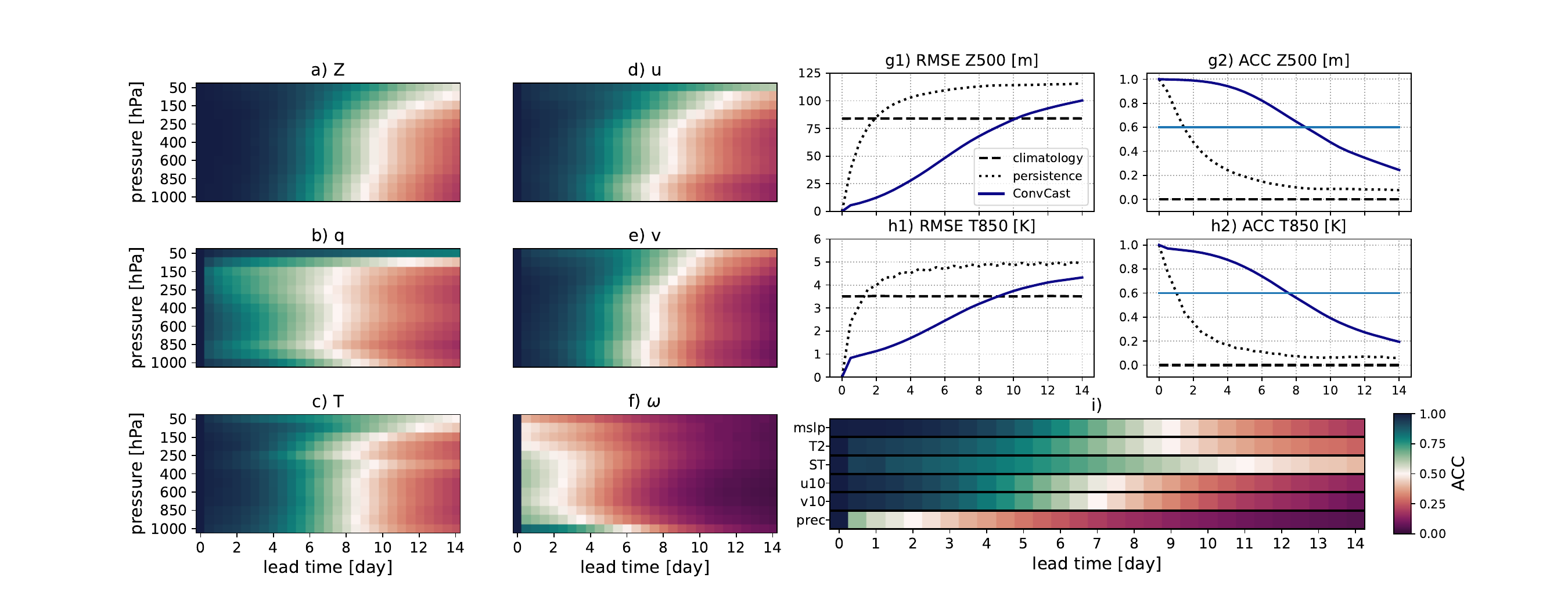}}
\caption{ACC at different lead times and pressure levels for a) $Z$, b) $q$, c) $T$, d) $u$, e) $v$, f) $\omega$. ACC for different lead times for i) mean sea level pressure ($mslp$), 2 m temperature ($T2$), surface temperature ($ST$), 10 m zonal wind ($u10$), 10 m meridional wind ($v10$) and precipitation ($prec$), and g1) globally averaged RMSE for 500 hPa geopotential ($Z500$), g2) ACC for $Z500$, h1) RMSE for 850 hPa temperature ($T850$) and h2) ACC for $T850$. The colour bar for anomaly correlation coefficient is shown at the bottom right. In panels g1, g2, h1, and h2, we plot the forecast skill of ConvCastNet prediction with solid, persistence with dotted, and climatology with a dashed line. \label{results_01_ACC_model_2002_2018}}
\end{figure*}

The forecast skill of the model is evaluated using root mean square error (RMSE) and anomaly correlation coefficient (ACC) for lead times up to 14 days. Forecasts are initialised from the test dataset, selecting every fifth day and alternating between 00 and 12 UTC initialisation times. Figure~\ref{results_01_ACC_model_2002_2018} shows the average ACC for all output variables (Table~\ref{tab: data}) and average RMSE for 500 hPa geopotential height ($Z500$) and 850 hPa temperature ($T850$). ACC is generally higher for $Z$, $T$, $u$, $v$, and $ST$ and significantly lower for $\omega$ and precipitation, variables that are more difficult to predict due to their high spatio-temporal variability, dominance of small-scale spectral components, and the model's limited resolution. Variables like $Z$, $T$, $T2m$, $mslp$, and $ST$ are more predictable partly because their power spectra are dominated by large-scale features, making them more compatible with ConvCastNet's coarse resolution. 

All variables except $\omega$ show shorter decorrelation time scales in the troposphere than in the stratosphere, consistent with \citet{Domeisen_Predictability_of_the_stratosphere}, where geostrophic, slowly evolving flow leads to more persistent patterns and greater predictability.

Most pressure level variables, except $T$, exhibit a discontinuity in ACC at 850 hPa, likely due to distinct boundary layer dynamics, and the fact that a large portion of the ERA5 training data at this level is subsurface. ConvCastNet achieves ACC > 0.6 up to 8.5 days for $Z500$ and 7.5 days for $T850$ (Figure~\ref{results_01_ACC_model_2002_2018} g2, h2). RMSE results show ConvCastNet outperforming persistence at all 14 lead days, and the 1979-2014 climatology for up to 10 days in $Z500$ and 9 days in $T850$.

\subsection{Spatial error distribution}\label{subsec: Spatial error spread}

We also assess forecast skill at individual model grid points by averaging errors over the test dataset, using the same forecast initialisation times as in Section~\ref{subsec: Forecast skill}. We compute both absolute errors, defined as the absolute difference between the prediction and the ERA5 reanalysis, and normalised errors, obtained by dividing the absolute errors by the local natural variability. This variability is defined as the standard deviation of the ERA5 reanalysis at the corresponding date and time, calculated over a $\pm$5-day window across the years 1979 to 2014. The normalised error reflects how the forecast error compares to typical weather variability at a given location and time of year. 

\begin{figure*}
\centerline{\includegraphics[width=1. \textwidth]{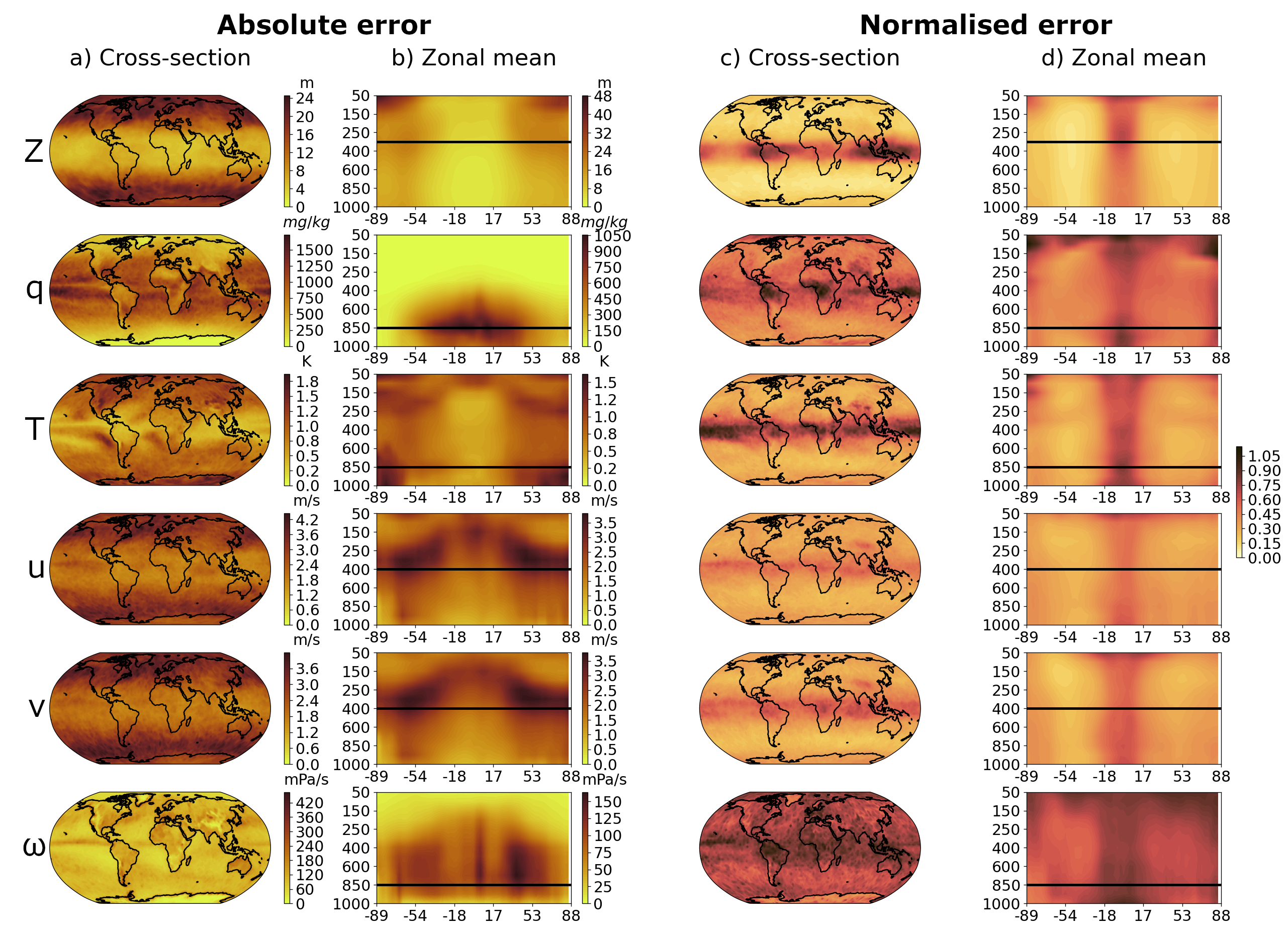}}
\caption{Average absolute (columns a) and b) ) and normalised (columns c) and d)) errors for predictions at lead day 2. Different rows show errors for $Z$, $q$, $T$, $u$, $v$ and $\omega$. Each absolute error figure is accompanied by its own colour bar. The unified colour bar for the normalised error is positioned on the right side of the image. A vertical cross-section of the zonal-mean error is plotted in columns b) and d). Columns a) and c) show a horizontal distribution of errors at the pressure level denoted by a corresponding black bar in columns c) and d).}\label{spatial_errors_time_step_4}
\end{figure*}

The largest absolute errors in geopotential height ($Z$) occur in the mid-to-high latitudes, peaking in the polar stratosphere above approximately 150 hPa (Figure~\ref{spatial_errors_time_step_4}a,b). Zonal and meridional winds ($u$ and $v$) exhibit the highest errors near the tropopause and in the lower mid-latitude troposphere, approximately in the regions of strong subtropical and mid-latitude westerly jets. Temperature errors are pronounced at low levels in the mid-latitudes and in the polar areas, as well as above the tropopause. Specific humidity ($q$) shows the largest errors in tropical and subtropical regions with high moisture content, especially over oceans. Although humidity decreases exponentially with height, the maximum error occurs near 850 hPa and not at the surface, and then declines sharply with increasing altitude. The errors in $\omega$ peak in regions of intense cyclogenesis (e.g.~in the Northern Atlantic and Northern Pacific), the inter-tropical convergence zone (ITCZ) and broadly over continental areas. Across all variables, regions with higher natural variability tend to exhibit larger absolute errors (not shown).

Normalised errors (Figure~\ref{spatial_errors_time_step_4}c,d) reveal different spatiotemporal distributions. At lead day 2, increased errors are primarily observed in the tropics and lower stratosphere. Since our model only partially represents the stratosphere, with the model top at 50 hPa level, rapid error growth at this level might be expected. Additionally, ML models often struggle to capture stratospheric dynamics as well as the physics-based models \citep[e.g.][]{GraphCast}. In contrast, the rapid error growth in the tropics is consistent with behaviour seen in physical models. The distribution of errors also evolves over time, with long-range normalised errors exhibiting a more uniform spatial pattern (see \ref{App: LongRangeSpatialErrorDistribution}). We speculate that the forecast skill could potentially be improved by applying date and time-specific grid-point standardisation, which would make small deviations more impactful in the loss function. 

\section{Methodology}\label{sec: Methodology}

In this study, we use the grid-point relaxation method to assess the influence of different parts of the atmosphere on model predictions, and apply gradient-based sensitivity analysis to investigate the dependence of forecast errors on initial conditions.

\subsection{Grid-point relaxation technique}

In traditional physics-based models, relaxation (nudging) typically involves adding a term of the form $-(1/\tau) (X - X_{\mathrm{ref}})$ to the model tendency equations (in either grid-point or spectral space) to relax the state $X$ towards a reference value $X_{\mathrm{ref}}$, with the parameter $\tau$ controlling the strength of the nudging. Variants such as spectral nudging have also been applied recently in hybrid physics-ML models \citep{husain2024leveragingdatadrivenweathermodels}. In contrast, we implement relaxation in our DL framework by directly forcing the model fields (not the tendencies). Specifically, we introduce a relaxation weight, $\Lambda$, and compute the relaxed state as a weighted linear combination of the value predicted by the model and the ground truth value based on ERA5 reanalysis: 
\begin{equation}\label{eq: weighted_prediction}
\begin{aligned}
   X(\mathbf{r}, t+\Delta t) &= \Lambda (\mathbf{r}) \cdot X^f(\mathbf{r}, t+\Delta t) + \\ &+ \left(1-\Lambda(\mathbf{r})\right) \cdot X^t(\mathbf{r}, t+\Delta t)\, .
\end{aligned}
\end{equation}
Here, $X$ denotes a prognostic variable, $X^t$ is the ERA5 ground truth, and $X^f(t+\Delta t)=\left(\mathrm{\mathcal{S}^{-1} \circ \mathcal{M}}_{t \rightarrow t + \Delta t} \circ \mathcal{S}\right)(X(t))$ represents the model forecast from time $t$ to $t+\Delta t$. The operator $\mathcal{S}$ denotes standardisation following Equation~\ref{eq: standardisation} and $\mathcal{S}^{-1}$ is its inverse (destandardisation). The position vector is denoted as $\mathbf{r}=(\lambda, \varphi, p)$. When $\Lambda=1$, the prediction is entirely based on the model output; when $\Lambda=0$, the forecast is replaced by ERA5 ground truth at the corresponding lead time. We define the relaxation weight as a function of pressure level and latitude as
\begin{equation}\label{eq: vertical}
\Lambda(p) = \pm \frac{1}{2}\tanh{\frac{p-p_0}{\Delta p}} + \frac{1}{2} \, ,
\end{equation}
and
\begin{equation}\label{eq: latitudinal}
\Lambda(\varphi) = \frac{1}{2} \tanh{\frac{\varphi - \varphi_{\mathrm{max}}}{\Delta \varphi}} - \frac{1}{2} \tanh{\frac{\varphi - \varphi_{\mathrm{min}}}{\Delta \varphi}} + 1 \, .
\end{equation}
The tangens hyperbolicus functions are applied to provide a smooth transition from the relaxation zone to the freely evolving region with $\Delta p$ and $\Delta \varphi$ determining the width of the transition area in vertical and meridional directions, and $p_0$, $\varphi_\mathrm{min}$ and $\varphi_\mathrm{max}$ the locations of transition, where $\Lambda=0.5$. 

In this study, we investigate the impact of applying relaxation in three regions with observed rapid error growth (see Sections~\ref{subsec: Forecast skill} and \ref{subsec: Spatial error spread}): the tropical troposphere, the global boundary layer, and the stratosphere. We further evaluate the resulting influence on mid-latitude weather predictability in the medium range.

The equatorial region ($\varphi_{\mathrm{min}} = -19.5 \degree$S, $\varphi_{\mathrm{max}} = 19.5 \degree$N, $\Delta \varphi=4.5$ $\degree$ in Eq.~\ref{eq: latitudinal}) was selected based on the observed rapid error growth (see Fig.~\ref{spatial_errors_time_step_4}) as well as results of previous studies showing the error growth originating from the small subsynoptic scales in the Tropics \citep{Zagar01012017} and spreading to mid-latitude synoptic scales \citep{Zagar2017}. Here we relax all prognostic variables, including $mslp$ and precipitation. 

We also evaluate the influence of surface and boundary layer processes on forecast skill in the upper troposphere and stratosphere. Here, the relaxation was performed using $p_0=850$ hPa and $\Delta p = 100$ hPa and a negative sign in Eq.~\ref{eq: vertical}. The relaxation zone was determined based on the aforementioned observed discontinuity in forecast skill (see Figure~\ref{results_01_ACC_model_2002_2018}). Relaxation is also applied in the fields of $mslp$, surface and 2 m temperature, and 10 m winds. Precipitation is not relaxed, as it is mostly formed in the lower to mid-troposphere above the boundary layer. 

The influence of an accurately represented stratosphere on tropospheric forecast skill has been explored in several studies \citep{Domeisen_Predictability_of_the_stratosphere, Domeisen2020, Kautz2020}. We assess this effect in a DL weather model by applying relaxation only above the dynamically determined thermal tropopause, using $\Delta p = 100$ hPa and a positive sign in Eq.~\ref{eq: vertical} to define the vertical transition. The tropopause was identified by first interpolating the temperature profile above each grid point to pressure levels using a cubic spline interpolator. We then computed the derivative of the interpolated temperature to locate extrema, selected the level corresponding to the expected tropopause, and applied a Gaussian filter with a standard deviation of 2 across horizontal dimensions to smooth the tropopause pressure field and suppress outliers.

We assessed the impact of relaxation by comparing freely evolving model predictions with relaxed forecasts, testing for statistically significant changes in absolute errors (Eq.~\ref{eq: error}) using the Mann-Whitney U test \citep{Mann-Whitney-U}. A 95\% confidence level was used to determine statistical significance. The tests were performed separately on vertically and zonally averaged absolute errors of the standardised fields, using forecasts from the test dataset initialised every five days, alternating between the 00 and 12 UTC. 

Additionally, we calculate the relative error change, defined as 
\begin{equation}\label{eq: relative_error_change}
 \mathrm{relative\hspace{1mm}error\hspace{1mm}\mathrm{change}} = \frac{\mathrm{error\hspace{1mm}with\hspace{1mm}relaxation}}{\mathrm{error\hspace{1mm}without\hspace{1mm}relaxation}} - 1 \, ,
\end{equation}
where "error with relaxation" and "error without relaxation" refer to the vertically or zonally averaged absolute errors of the prediction at a given lead time, with and without relaxation, respectively.

\subsection{Error sensitivity}

We analyse the sensitivity of the model's prediction errors to the initial conditions by computing the gradient of an error function $E$, defined as the sum of absolute differences between standardised reanalyses and model predictions over a set $\mathcal{V}$ of atmospheric variables within a specified domain $\mathcal{D}$:
\begin{align}\label{eq: error}
 E &= \sum_{v\in\mathcal{V}} \sum_{\mathbf{r} \in \mathcal{D}} \left| \hspace{1mm} \mathbf{S}_v^t(\mathbf{r}, t) - \mathbf{S}_v^f(\mathbf{r}, t) \right| \\
&= \sum_{v\in\mathcal{V}} \sum_{\mathbf{r} \in \mathcal{D}} \left| \hspace{1mm} \mathbf{S}_v^t(\mathbf{r}, t) - \left[\boldsymbol{\mathrm{\mathcal{M}}}_{0 \rightarrow t}(\mathbf{S}(t=0))\right]_v(\mathbf{r})  \hspace{1mm} \right|,
\end{align}
where $\mathcal{D}$ represents a closed volume of atmosphere for which the sensitivity is calculated, $\mathbf{r}=(\lambda,\varphi,p)$ is a position vector, $\mathbf{S}^t=\mathcal{S}\left(\mathbf{X}^t\right)$ is the standardised ERA5 ground truth tensor, and $\mathbf{S}^f=\mathcal{S}(\mathbf{X}^f)$ is the predicted tensor in standardised form. Using the DL model's auto-differentiability, we can compute the sensitivity of the error to the initial condition:
\begin{equation}\label{eq: grad_S E}
    \frac{\partial E} {\partial \mathbf{S}(0)} = -\sum_{v\in\mathcal{V}} \sum_{\mathbf{r} \in \mathcal{D}} \mathrm{sgn}\left(\mathbf{S}_v^t(\mathbf{r}, t) - \left[\boldsymbol{\mathrm{\mathcal{M}}}_{0 \rightarrow t}\left(\mathbf{S}(0)\right)\right]_v(\mathbf{r})\right) \left[\nabla_{\mathbf{S}(0)}\boldsymbol{\mathrm{\mathcal{M}}}_{0 \rightarrow t}(\mathbf{S}(0))\right]_v(\mathbf{r}) \,,
\end{equation}
where the term
\begin{equation*}
    \nabla_{\mathbf{S}(0)}\boldsymbol{\mathrm{\mathcal{M}}}_{0 \rightarrow t}\left(\mathbf{S}\left(0\right)\right)=\frac{\partial \boldsymbol{\mathrm{\mathcal{M}}}_{0 \rightarrow t}(\mathbf{S}(0))} {\partial \mathbf{S}(0)}
\end{equation*}
represents the sensitivity of the model output at time $t$ to changes in the standardised initial condition $\mathbf{S}(0)$ and is computed using backpropagation. Notably, the gradient of the standardised error with respect to standardised inputs is equivalent to the gradient of the non-standardised error with respect to non-standardised inputs (see \ref{App: Non-standardised sensitivity} for details). The sign function in Eq.~\ref{eq: grad_S E} is defined as
\begin{equation*}
\mathrm{sgn}(x) =
    \begin{cases}
      1, & \text{if}\ x \geq 0 \\
      -1, & \text{if}\ x < 0
    \end{cases}.    
\end{equation*}
Except for the sign, the resulting gradient is the same as the gradient of the model's prediction.

We have also performed the analysis of the average magnitude of error sensitivity (AMES), i.e. the magnitude of error sensitivity, averaged over several years, as $\overline{\left|\partial E\, /\, \partial \mathbf{S}(0)\right|}$.

\section{Results}\label{sec: Results}

\subsection{Grid-point relaxation}\label{sec: Grid-point relaxation}

\begin{figure*}
\centerline{\includegraphics[width=1. \textwidth]{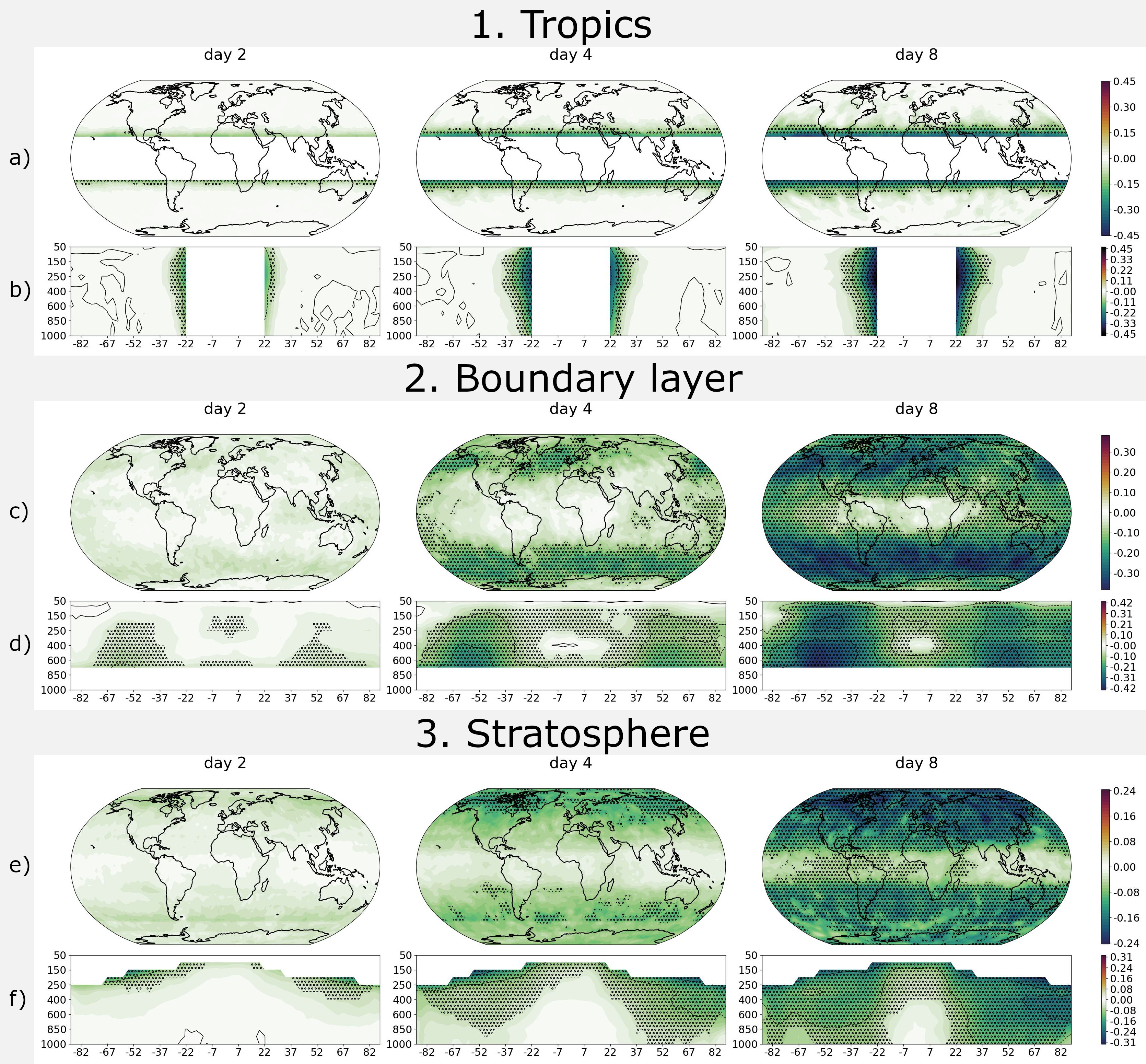}}
\caption{Panels 1-3 show the relative temperature error change (see Eq.~\ref{eq: relative_error_change}) for relaxation in the tropics (case 1), boundary layer relaxation (case 2), and stratosphere relaxation (case 3), respectively. Panels (a, c, e) display the relative change in vertically-averaged temperature error and panels (b, d, f) show the relative change of zonal-mean temperature error. A negative value indicates that the average error has decreased, implying improved forecast accuracy. The contour spacing in (b, d, f) is 0.1. Statistically significant changes are marked with hatched black circles. Relaxation zones, where $\Lambda\geq0.5$, are masked in white.}\label{relaxation_time_steps}
\end{figure*}

\begin{figure*}
\centerline{\includegraphics[width=1.\textwidth]{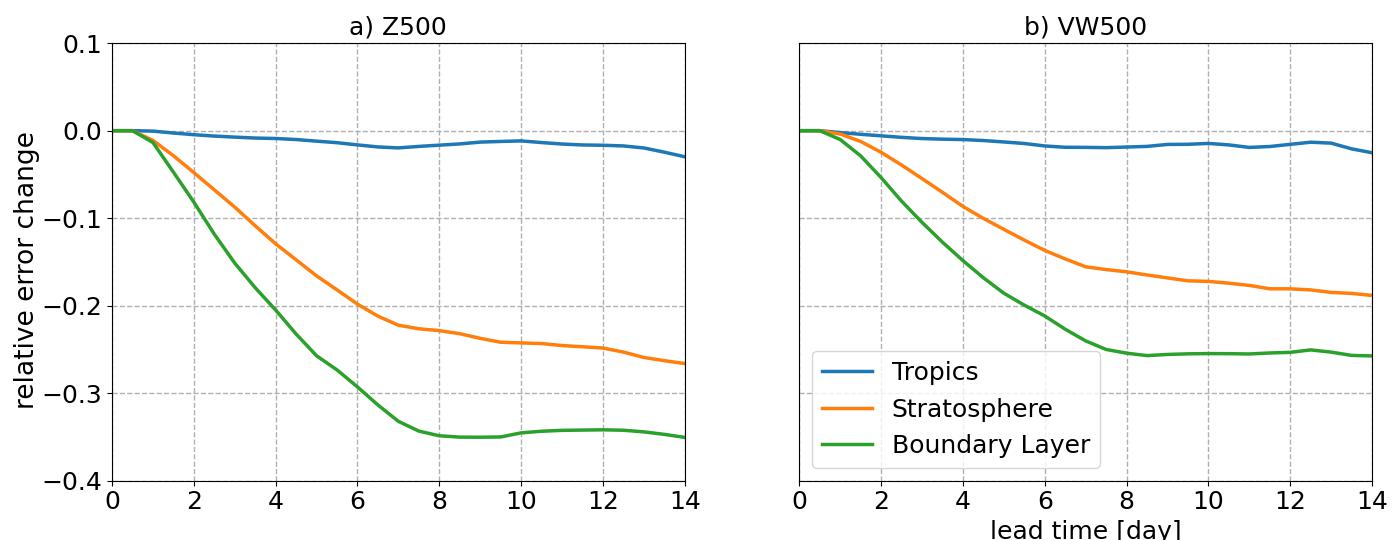}}
\caption{Relative error change (Eq.~\ref{eq: relative_error_change}) of 500 hPa geopotential height, $Z500$ (a), and 500 hPa vector wind, $VW500$ (b) for the Northern Hemisphere extratropics. The error is determined using root mean square error (RMSE), where squared errors for a given lead time are averaged over $\varphi \in [19.5\degree, 90\degree]$ and all forecast initialisations. \label{relaxation_relative_error_change}}
\end{figure*}

We first examine the relative error change in the temperature predictions for all three cases, described in Section~\ref{sec: Methodology}. In the tropical case (Fig.~\ref{relaxation_time_steps}, case 1), a statistically significant reduction in relative error is observed primarily near the tropical relaxation domain, with the improvement gradually extending in the meridional direction. After 8 days, the influence of the relaxation extends approximately 10 - 15$\degree$ north and south of the $\Lambda = 0.5$ relaxation threshold. The relaxation influence spreads poleward unevenly, with the most pronounced impact observed between 100 and 500 hPa, as observed from the vertical profile of the zonal-mean error change in Fig.~\ref{relaxation_time_steps}b. Tropical relaxation doesn't significantly impact forecast skill in the mid-latitudes and polar regions, in line with \citep{Wang2025}. In some localised areas, forecast skill improves by up to 5\%, but the difference is not statistically significant. This pattern is also evident in Fig.~\ref{relaxation_relative_error_change}, which shows the relative RMSE change in the Northern Hemisphere extratropics. Only a slight improvement in forecast skill over lead time is obtained when relaxation is applied in the tropics. Based on the results in \citep{Wang2025}, increasing the model resolution would not significantly alter the results, as the small-wavelength (high wavenumber) features are more likely to remain trapped in the equatorial waveguide.

In the second case, i.e. the boundary layer relaxation (Fig.~\ref{relaxation_time_steps}, case 2), the influence of relaxation propagates primarily in the vertical direction. It gradually affects the mid-latitude jet stream region and subsequently extends into the tropical atmosphere. The impact is most pronounced above mid-latitude oceanic regions with strong baroclinic activity. The spread of influence into the tropical upper-troposphere reflects the generally strong surface and upper-troposphere coupling in the convective regions \citep{Gill1980}, while the impact on the tropical mid-troposphere is limited.

In stratospheric relaxation (Fig.~\ref{relaxation_time_steps}, case 3), the initial stratospheric impact descends into the extratropical troposphere in the medium-range. This delayed impact is in line with \citet{Roff2011}, who also demonstrated the delayed tropospheric impact of improved representation of the stratosphere in physics-based models. Improvements in mid-latitude stratospheric forecasts are also observed with boundary layer relaxation, indicating strong two-way troposphere-stratosphere coupling. The downward propagating impact of stratospheric relaxation is less pronounced than the upward propagating impact of the boundary layer relaxation (compare relative error change in Fig.~\ref{relaxation_relative_error_change}); however, this difference is strongly sensitive to the applied tropopause level in stratospheric relaxation.

\begin{figure*}
\centerline{\includegraphics[width=1. \textwidth]{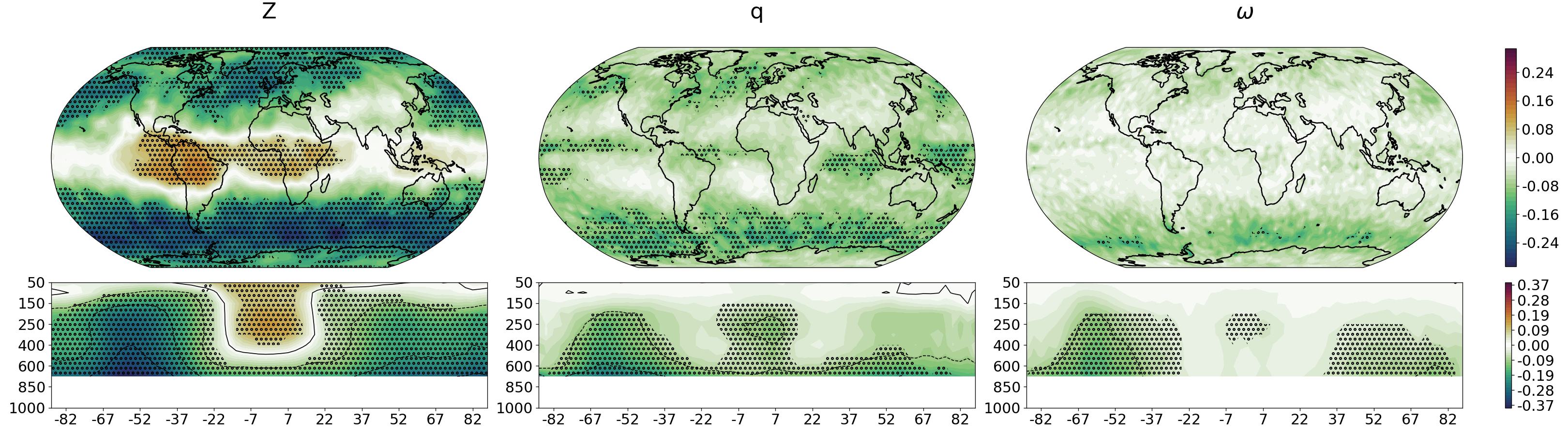}}
\caption{Same as Figure~\ref{relaxation_time_steps}, case 2, lead time of 4 days, but for geopotential height $Z$, specific humidity $q$ and vertical velocity $\omega$.} \label{relaxation_variables}
\end{figure*}

\begin{figure*}
\centerline{\includegraphics[width=1. \textwidth]{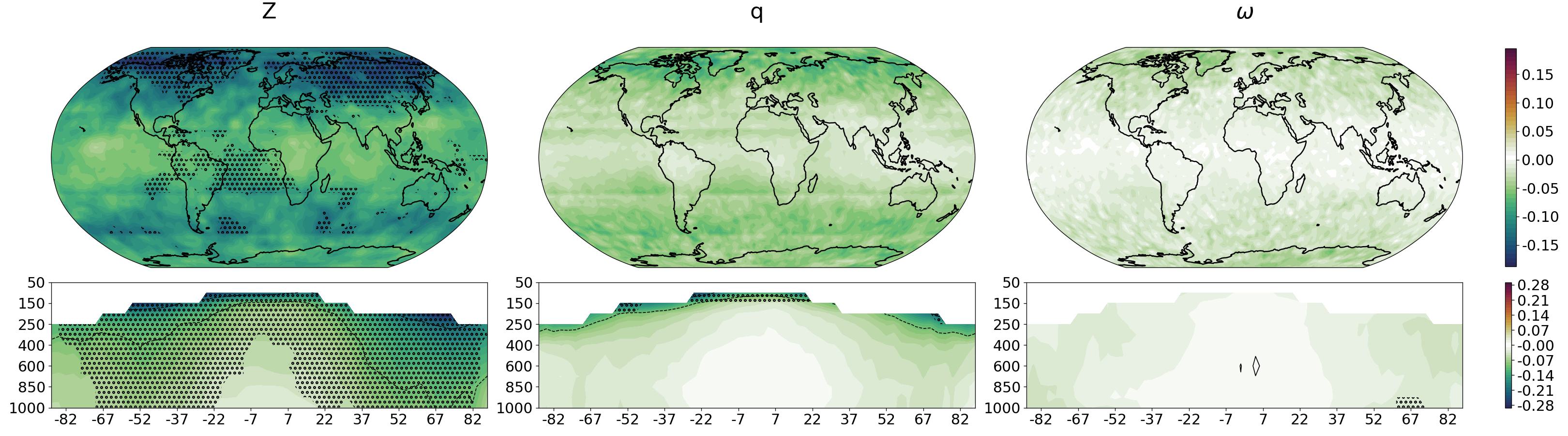}}
\caption{Same as Figure~\ref{relaxation_variables}, but for relaxation in the stratosphere. \label{relaxation_variables_stratosphere}}
\end{figure*}

The effects of relaxation vary across different variables. For example, in Figure~\ref{relaxation_variables}, we compare the relative error changes in geopotential height $Z$, specific moisture $q$, and vertical velocity $\omega$ under boundary layer relaxation at lead day 4. The strongest improvements are seen in $Z$ and $T$ (not shown), while the impact on $q$ and $\omega$ is weaker. These improvements are particularly evident over mid-latitude oceans, especially in regions of strong baroclinicity such as the North Atlantic, North Pacific and the Southern Ocean. Additionally, improved forecasts for $q$ and $\omega$ are seen in the intertropical convergence zone (ITCZ), with similar enhancements observed for the winds (not shown). In most regions and variables, relaxation improves the forecasts. However, this is not the case for the geopotential prediction in the tropical mid- and upper-troposphere and global stratosphere, where relaxation appears to degrade geopotential forecasts, most notably in a belt stretching from central Africa to the Eastern Pacific. The same analysis is also performed for stratospheric relaxation (Fig.~\ref{relaxation_variables_stratosphere}). In this case, the differences between different variables are more pronounced. While temperature $T$ and geopotential height $Z$ show clear improvements, specific humidity $q$ and vertical velocity $\omega$ exhibit only a slight and insignificant gain.

\subsection{Error sensitivity}

To complement the relaxation analysis, we perform a case study of Hurricane Ian’s forecast error sensitivity to initial conditions and compute the average magnitude of error sensitivity (AMES) for selected regions using the samples from the test dataset.

\subsubsection{Hurricane Ian}

\begin{figure*}
\centerline{\includegraphics[width=1. \textwidth]{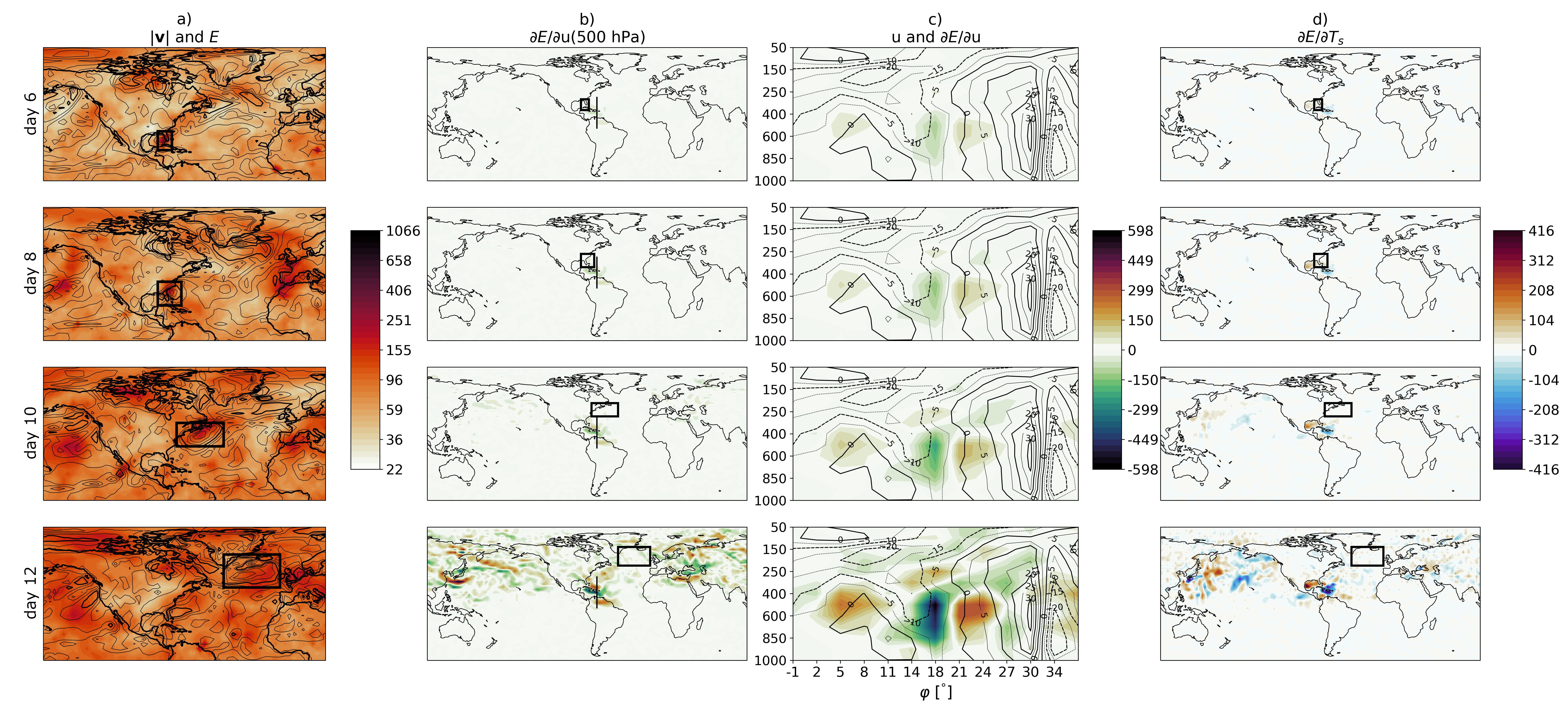}}
\caption{Column a) shows the 10 m horizontal wind speed (contours at 2.5 m/s intervals, with bold contours at 10, 20, and 30 m/s) and the absolute error summed over all variables and pressure levels (colours). Column b) presents the sensitivity of the absolute error within the black box to the 500 hPa zonal wind (u). Column c) displays the vertical cross-section of u (contours) and its associated error sensitivity (colours) at $\lambda = -69^\circ$ and $\varphi \in [-1.5^\circ, 40.5^\circ]$ (corresponding to the black lines in column b)). Column d) shows the error sensitivity to surface temperature $T_s$. Rows represent different forecast lead times: day 6, day 8, day 10, and day 12. The black boxes in panels a) and b) indicate the error calculation domain $\mathcal{D}$. Columns b) and c) share the colourbar. Colourbar for a) is in log scale.} \label{fig: Error_Sensitivity_TC_Ian}
\end{figure*}

We examine the tropical-to-extratropical transition of Hurricane Ian by evaluating the error sensitivity to initial conditions at lead times 6, 8, 10 and 12 days. The domain $\mathcal{D}$ (see Eq.~\ref{eq: error}) is selected to include the cyclone and the surrounding region with the elevated forecast error. The forecast is initialised on September 23rd 2022, at 00 UTC. Approximately at that time, Hurricane Ian formed as a tropical depression in the central Caribbean \citep[see][]{NWS_HurricaneIan_2022}. Figure~\ref{fig: Error_Sensitivity_TC_Ian}a shows that the absolute forecast errors associated with Hurricane Ian remain tightly concentrated around the hurricane during the first eight days. Beyond this period, the predicted storm track diverges from the observed trajectory as the westerly jet steers the system toward the North Atlantic. 

During the first 8 days, columns b) and c) in Fig.~\ref{fig: Error_Sensitivity_TC_Ian} show that the error sensitivity to the zonal wind is largely confined to the mid-troposphere over the Caribbean. After 10 and 12 days, error sensitivity spreads upstream to the Northern and Western Pacific, Asia and Europe. The error sensitivity diminishes towards the equator and the pole and, for the most part, does not extend across the equator into the Southern Hemisphere. The vertical cross-section reveals both amplification and upward spreading of sensitivity to the initial mid-tropospheric zonal wind, extending into the stratosphere.

The distribution of error-sensitivity to zonal wind in the Caribbean persists throughout the 12-day evolution of Hurricane Ian, with sensitivity increasing over time. This behaviour is expected, as small perturbations in the initial trajectory can lead to large positional differences at longer lead times. South of the Caribbean islands (at $\varphi=18^\circ$) at the 500 hPa level, where $\partial E/\partial u < 0$, a small positive zonal wind increment would reduce the forecast error (Fig.~\ref{fig: Error_Sensitivity_TC_Ian}c)). Conversely, north of the Caribbean islands (at $\varphi=22^\circ$), where $\partial E/\partial u > 0$, a negative increment is needed. Qualitatively, this implies that the westerly subtropical jet stream should be slightly shifted southward in the initial conditions to improve the forecast of Hurricane Ian.

The error exhibits only a limited sensitivity to Hurricane Fiona, centred downstream of Ian at around $\varphi=32^\circ$N at the initialisation time (see wind contours in Fig.~\ref{fig: Error_Sensitivity_TC_Ian}c). This finding aligns with \citet{Quinting2016_TC_RWP}, who demonstrated that tropical cyclones primarily influence the downstream Rossby wave packets' amplitude and frequency. Consequently, the limited error sensitivity associated with Hurricane Fiona is expected as Hurricane Ian is situated upstream. Forecast error sensitivity to surface temperature displays a similar spatial pattern: short-term influence is concentrated over the Caribbean and Gulf of Mexico, while long-term influence propagates upstream to the Northern and Western Pacific. The forecast error for Hurricane Ian is primarily sensitive to sea surface temperature (SST), with considerably lower sensitivity to land and ice surface temperatures. This heightened sensitivity to SST is expected, as warmer ocean surfaces promote faster tropical cyclone formation and more rapid intensification \citep{Ren2014_TC_SST}.

\subsubsection{Climatological error sensitivity}

To generalise the sensitivity findings and make sense of the results from the tropical relaxation experiment in Section~\ref{sec: Grid-point relaxation}, we compute "climatological" error sensitivity across the test dataset (2020-2022) for the three predefined latitude bands:
\begin{enumerate}
    \item $\mathcal{D}_A = \{(\varphi, \lambda, p); \varphi \in [-4.5,4.5]\degree\}$,
    \item $\mathcal{D}_B = \{(\varphi, \lambda, p); \varphi \in [28.5,34.5]\degree\}$,
    \item $\mathcal{D}_C = \{(\varphi, \lambda, p); \varphi \in [40.5,49.5]\degree\}$,
\end{enumerate}
where in all cases, $\lambda \in [0,360)\degree$ and $p\in\{1000, 925, 850, 700, 600\}$ hPa. The absolute error calculation includes precipitation, mean sea-level pressure ($mslp$), surface temperature ($T_s$), 2 m temperature ($T_2$), and 10 m wind components. 

\begin{figure*}
\centerline{\includegraphics[width=1. \textwidth]{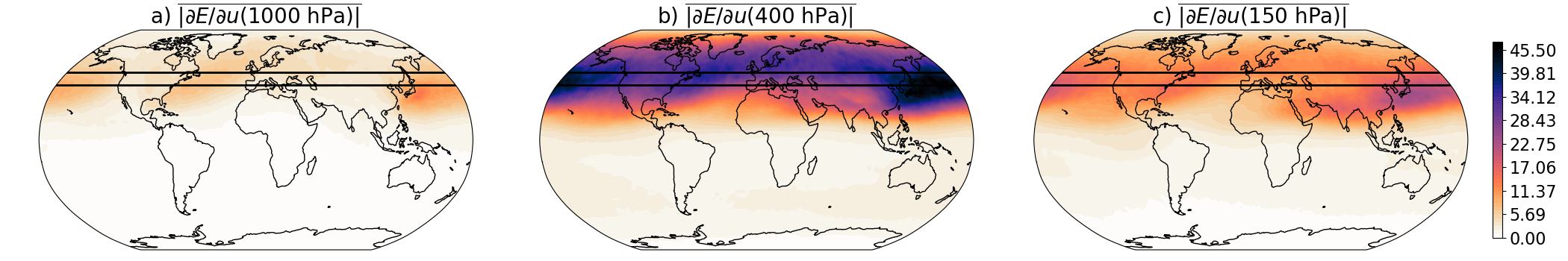}}
\caption{Climatological sensitivity of forecast error on lead day 8, computed within domain $\mathcal{D}_C$ (the volume enclosed by the black lines), to zonal wind ($u$), based on the 2020–2022 test period. Panel (a) shows the sensitivity at 1000 hPa, (b) at 400 hPa, and (c) at 150 hPa.} \label{fig: Error_Sensitivity_pressure_levels}
\end{figure*}

First, we compute the AMES in domain $\mathcal{D}_C$ to the initial zonal wind at different pressure levels (Fig.~\ref{fig: Error_Sensitivity_pressure_levels}). Both the magnitude and spatial distribution of sensitivity vary notably with height. Sensitivity is predominantly confined to the Northern Hemisphere. Two regions of local maximum sensitivity are located in Asia and North America's highly baroclinic eastern flanks. The sensitivity diminishes towards the West coasts of North America and Europe and is less significant over the continents. These results also suggest which regions should be particularly well initialised (and also observed) in order to minimise the medium-range forecast error in this latitude belt.

\begin{figure*}
\centerline{\includegraphics[width=1. \textwidth]{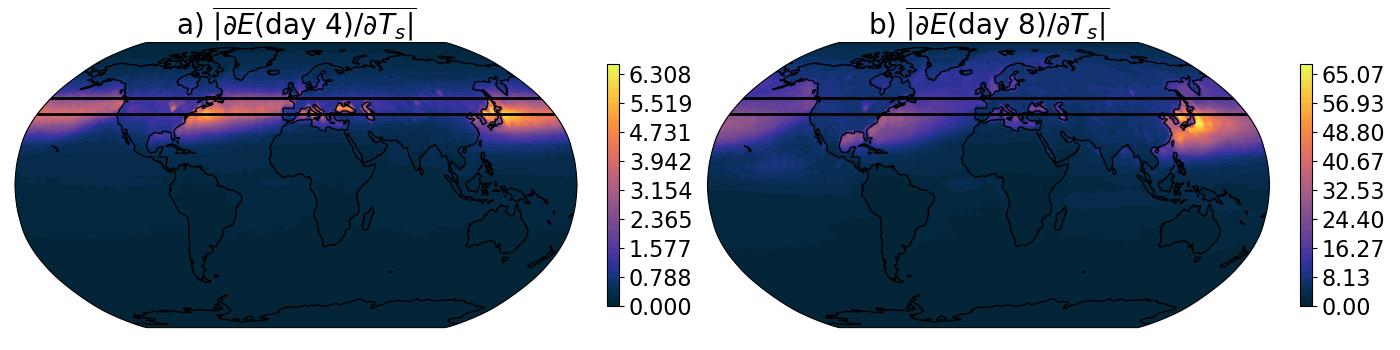}}
\caption{Average magnitude of error sensitivity to surface temperature $T_s$ in domain $\mathcal{D}_C$ (volume enclosed by the black lines) for a) 4 and b) 8-day forecasts.}\label{fig: Error_Sensitivity_time_dependence}
\end{figure*}

Figure~\ref{fig: Error_Sensitivity_time_dependence} presents the AMES for the same domain at two different forecast lead times to initial surface temperature $T_s$. The forecast shows the greatest sensitivity to $T_s$ over ocean surfaces —i.e., sea surface temperature (SST)—while sensitivity over land and ice surfaces is limited. The sensitivity is greatest in regions with strong meridional SST gradients, particularly where warm ocean currents encounter cooler waters along the eastern boundaries of continents. In the 4-day forecast, we see the region of sensitivity confined mostly to the areas inside and slightly to the south of the $\mathcal{D}_C$ domain. The southernmost sensitive regions reach down to the Gulf of Mexico and the East China Sea. The AMES of the 8-day prediction increases in magnitude and spreads further to the North and South. We see an increased sensitivity to the Arctic seas and an expansion of the sensitivity to the Gulf of Mexico, the Eastern Pacific and the South China Sea. 

\begin{figure*}
\centerline{\includegraphics[width=1. \textwidth]{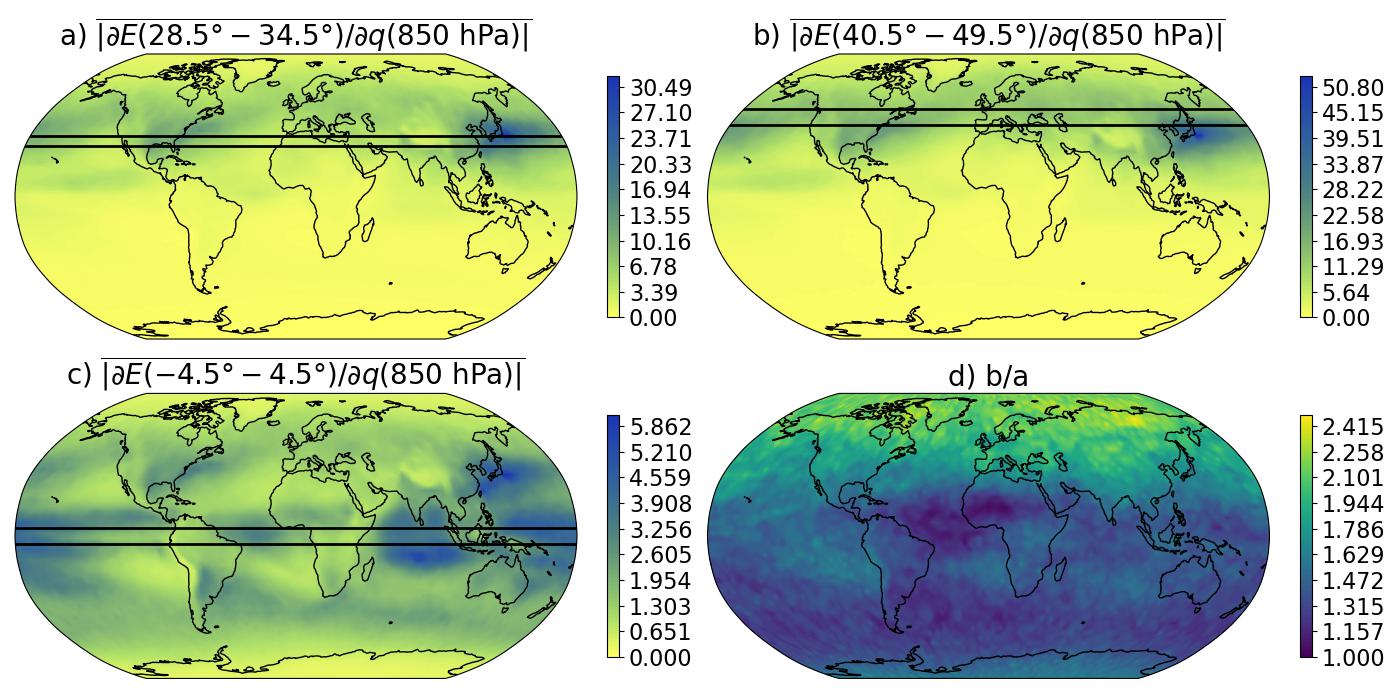}}
\caption{Same as Figure~\ref{fig: Error_Sensitivity_pressure_levels}, but for absolute error sensitivity to the 850 hPa specific humidity $q$ with domains a) $\mathcal{D}_B$, b) $\mathcal{D}_C$ and c) $\mathcal{D}_A$. Panel d) shows the sensitivity from $\mathcal{D}_C$ (panel b) divided by the sensitivity from $\mathcal{D}_B$ (panel a). \label{fig: Error_Sensitivity_latitude_dependence}}
\end{figure*}

In Fig.~\ref{fig: Error_Sensitivity_latitude_dependence}, we compare AMES to specific humidity $q$ at 850 hPa across different error domains. Figures~\ref{fig: Error_Sensitivity_latitude_dependence}ab show considerable sensitivity to the Eastern flanks of the continents, the Central and Eastern North America area, the Ganges and Brahmaputra basins and other regions, including the Pacific ITCZ. The difference in distribution is illustrated in Figure~\ref{fig: Error_Sensitivity_latitude_dependence}d, which displays the sensitivity ratio of domain $\mathcal{D}_C$ to $\mathcal{D}_B$. While the domains differ in volume—precluding a direct quantitative comparison—a northward shift in sensitivity is nonetheless evident. In Fig.~\ref{fig: Error_Sensitivity_latitude_dependence}c, errors in domain $\mathcal{D}_A$ exhibit sensitivity to the tropical region as well as to the mid-latitudes of both the Northern and Southern Hemispheres. The highest sensitivity is found over the Indian Ocean, the western Pacific, and the lee sides of major mountain ranges.

Since AMES identifies regions in the input fields that have the greatest influence on forecast error, it serves as a valuable complement to the relaxation experiments discussed in Section~\ref{sec: Grid-point relaxation}. The sensitivity shown in Fig.~\ref{fig: Error_Sensitivity_latitude_dependence}a indicates that 8-day forecast errors in domain $\mathcal{D}_B$ are sensitive to the 850 hPa specific humidity $q$ within the equatorial region. This result aligns with the statistically significant 8-day impact reaching upwards to 34$\degree$ in the tropical relaxation experiment (Fig~\ref{relaxation_time_steps}, panel 1a). The fact that the domain $\mathcal{D}_C$ lies further from the tropics, in an area not significantly impacted by tropical relaxation, indicates a smaller sensitivity to equatorial regions. Computed sensitivity shows this is evidently not the case (see Figure~\ref{fig: Error_Sensitivity_latitude_dependence}b). A reduced tropical influence on medium-range forecasts in this region is therefore possibly a result of greater sensitivity to the mid-latitude and polar regions (Fig.~\ref{fig: Error_Sensitivity_latitude_dependence}d), which may make tropical influence negligible.

In addition, the 8-day error sensitivity to surface temperature in the mid-latitude domain $\mathcal{D}_C$ (Fig.~\ref{fig: Error_Sensitivity_time_dependence}b) appears relatively insensitive to tropical oceanic conditions. Instead, stronger sensitivity is observed over oceanic regions adjacent to the eastern boundaries of continents, suggesting that sea surface temperatures (SST) in these areas play a more prominent role in mid-latitude forecast error growth than those in the tropical oceans.

The variation in the meridional extent of sensitivity across pressure levels (Fig.~\ref{fig: Error_Sensitivity_pressure_levels}) may help explain the differing patterns of poleward propagation seen in the relaxation experiment (Fig.~\ref {relaxation_time_steps}, panel 1b). In particular, error reductions spread more rapidly poleward between 100 and 500 hPa, which corresponds to a broader latitudinal AMES at these levels.

\section{Conclusions and Outlook}\label{sec: Conclusions}

This study used ConvCastNet, a deep-learning-based global weather prediction model, to perform relaxation experiments and assess the sensitivity of forecast errors to initial conditions. This allows for the identification of the regions where the atmospheric processes require improved representation and initial conditions need better specification to extend the medium-range predictability.

An efficient design of ConvCastNet enables computationally inexpensive yet skilful forecasts, making it well-suited for analysing deep learning weather prediction model behaviour and forecast errors. Analogous to the traditional relaxation technique, we implemented a grid-point relaxation approach to analyse the regional influence on deep learning–based weather predictions. Three cases were examined, where predicted fields were nudged towards ERA5 ground truth: the tropics, boundary layer and stratosphere. Results show that tropical relaxation has limited influence on extratropical predictability. Conversely, boundary layer and stratospheric relaxation significantly impact mid-latitude predictability. The effects of the relaxation propagate vertically, with substantial improvements in stratospheric prediction observed after approximately eight days of boundary layer relaxation. This suggests a strong coupling between the troposphere and stratosphere—a phenomenon that has been the subject of extensive research \citep[see, e.g.][]{Polvani2004}. Deep learning models, with their ability to capture complex, nonlinear relationships efficiently, could offer valuable insights into this coupling. 

By analysing error sensitivity to initial conditions during Hurricane Ian’s tropical-to-extratropical transition, we demonstrate that hurricane forecast error sensitivity propagates upstream with increasing lead time, consistent with the downstream propagation direction of Rossby wave packets. We also find higher sensitivity to sea surface temperatures compared to land and ice surface temperatures, highlighting the importance of an accurate ocean–atmosphere coupling in weather prediction models. This result is expected and likely stems from an important role of sea surface temperature in determining the sensible and latent heat fluxes, e.g.~through evaporation, which fuels storms and releases latent heat into the atmosphere, thereby impacting atmospheric dynamics. 

Figure~\ref{fig: Error_Sensitivity_latitude_dependence} further illustrates a difference in average magnitude of error sensitivity between the tropics and extratropics, showing that the extratropics are barely impacted by the initial conditions in the tropics. The finding is consistent with our tropical relaxation experiments (Fig.~\ref{relaxation_time_steps}). Specifically, the tropical influence on mid-latitude weather prediction appears constrained in both magnitude and spatial extent, with the strongest impact originating from the tropical Pacific. 

The effectiveness of these methods is inherently tied to the capabilities of the underlying model, and several areas for improvement have been identified. Since gradient-based sensitivity analysis relies on differentiating forecast error with respect to input fields, an auto-differentiable model is essential. Our current architecture uses the LeakyRELU activation function, which introduces non-differentiability at $x=0$. Replacing it with a smooth alternative such as ELU or GELU could improve the stability and interpretability of gradient-based diagnostics. Similarly, the choice of error function affects the gradient behaviour: the absolute error (Eq.~\ref{eq: error}), used here, is not differentiable at zero and may be less suitable in regions with small errors. Using a differentiable loss function, such as mean squared error, could offer benefits in such cases.

Further benefits could be achieved by increasing model resolution. Higher horizontal resolution could improve the localisation of sensitivity and relaxation impacts, while finer vertical resolution could better capture vertical error propagation, especially at the tropopause, where vertical gradients are large. Increasing the model top could enhance the impact of the stratosphere relaxation. Since both relaxation and sensitivity techniques are model-dependent, future work should explore their application across different DL architectures to assess their generality and robustness. Beyond diagnostic purposes, these methods could be used for hypothesis testing in dynamical meteorology — for example, examining the role of soil moisture and other predictors in heatwave development \citep[e.g.][]{PredictabilityLimit}, stratosphere–troposphere coupling or the influence of teleconnections in subseasonal-to-seasonal predictions. It should also be noted that care is needed in the interpretation of the results. For example, since the stratosphere and troposphere are two-way coupled (both downward and upward), relaxation of the stratosphere to "truth" exaggerates the actual improvement that would be achieved only by improving the model's representation of the stratospheric processes, as the forecast errors from the troposphere would still propagate upwards. Similar considerations should be applied to other dynamical processes as well.

The approaches outlined here can also inform improvements in data assimilation and the development of the observing system. In particular, sensitivity analysis could support the design of targeted observing systems by identifying regions where additional measurements would yield the greatest forecast error reduction or where the reduction of forecast error would bring the greatest societal benefit \citep{TargetedObservations, Venutti2025}. Looking ahead, we plan to apply this methodology with a focus on Europe, aiming to identify key areas that influence the predictability of European weather, which could guide future developments in observation system design.

Overall, our findings suggest that grid-point relaxation and error sensitivity provide a valuable set of tools for improving model understanding, forecast accuracy, and potentially the design of observing systems. When used in combination, relaxation experiments and sensitivity analysis offer a powerful means to interpret, evaluate, and enhance the performance of data-driven weather prediction models. As deep learning continues to expand its role in atmospheric science, such interpretability tools could play a vital role in future weather forecasting systems.

\section*{Acknowledgements}
Uro\v s Perkan acknowledges funding from the Slovenian Research and Innovation Agency (ARIS) Programme P1-0188 and Grant MR-58115. \v Ziga Zaplotnik acknowledges funding by the European Union under the Destination Earth initiative. The authors would also like to acknowledge funding from Grant SN-ZRD/22-27/0510. The authors would like to acknowledge Bo\v stjan Melinc (University of Ljubljana, Faculty of Mathematics and Physics) for his careful reading of the initial manuscript and valuable comments.

\section*{Conflict of interest}
The authors declare no potential conflict of interest.

\section*{Data availability statement}

The experiments in this paper were performed using ERA5 data \citep{ERA5_PressureLevels, ERA5_SingleLevels}, which was downloaded from the Copernicus Climate Change Service (C3S) Climate Data Store (accessed October 5th 2023). The results contain modified Copernicus Climate Change Service information from 2023. Neither the European Commission nor ECMWF is responsible for any use that may be made of the Copernicus information or data it contains.

\section*{Orcid}
Uro\v s Perkan: \href{https://orcid.org/0009-0003-7453-8700}{https://orcid.org/0009-0003-7453-8700}\newline
\v Ziga Zaplotnik: \href{https://orcid.org/0000-0002-6012-0480}{https://orcid.org/0000-0002-6012-0480}\newline
Gregor Skok: \href{https://orcid.org/0000-0003-0265-1205}{https://orcid.org/0000-0003-0265-1205}


\bibliographystyle{abbrvnat}
\bibliography{references}


\appendix
\renewcommand{\thesection}{Appendix \Alph{section}} 

\titleformat{\section}
  {\normalfont\Large\bfseries}
  {Appendix \Alph{section}:}{1em}{}

\section{ConvCastNet Details}
\label{App: TechnicalDetails}
\vspace{12pt}

\begin{figure*}
\centerline{\includegraphics[width=0.6 \textwidth]{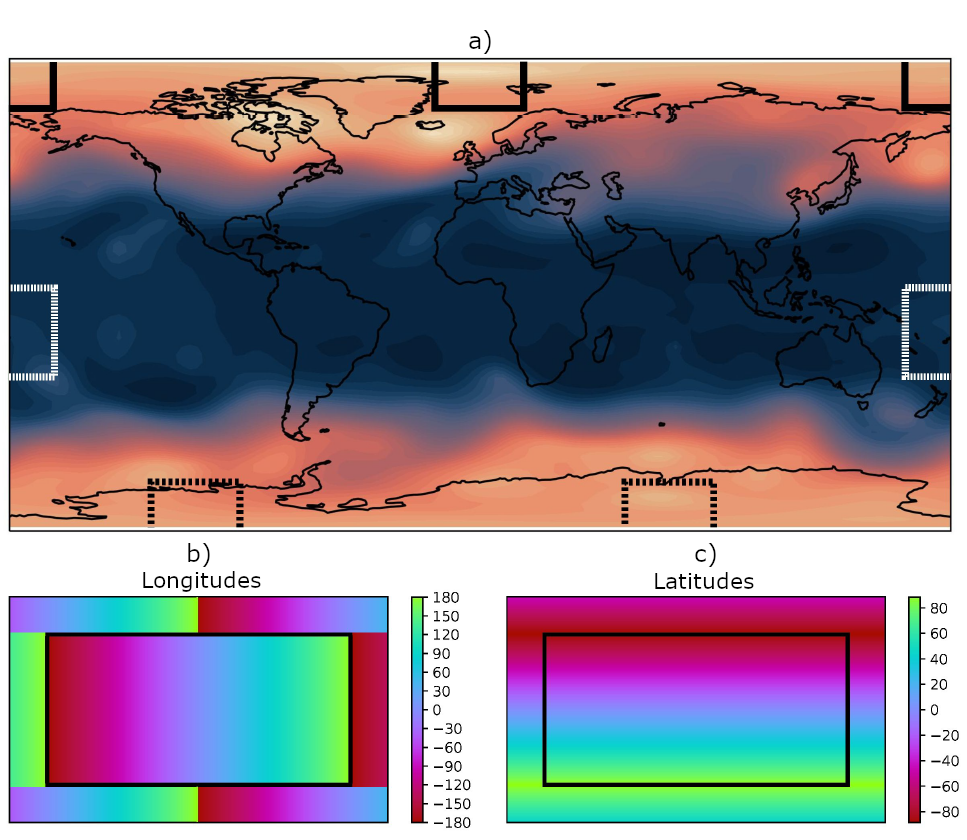}}
\caption{Figure (a) illustrates the spatial regions that convolutional kernels must cover in order to overlap a compact patch of ground on the sphere. The white dashed box highlights the desired kernel coverage near the antimeridian boundary. The black dashed box shows the synonymous kernel coverage near the south pole, away from the antimeridian, while the solid black box indicates the desired coverage near the north pole at the antimeridian. Figures (b) and (c) display the longitude and latitude arrangements of the padded tensors designed to meet these coverage requirements.\label{fig: SphericalPadding}}
\end{figure*}

ConvCastNet is a convolutional encoder-decoder neural network, based on the U-Net architecture. A short overview of the model is described in Section~\ref{sec: Model architecture}. The model is based on depthwise separable (DS) convolutions with a stride of one. All depthwise convolutions except for the first convolution in the first block use kernels of size $3 \times 3$. The first depthwise convolution, operating on the input tensor, uses a kernel size $7 \times 7$ which corresponds to a $21 \degree \times 21 \degree$ area in the physical input space in order for each kernel to be able to learn synoptic scale spatial patterns. The Leaky ReLU activation function was selected based on empirical testing, where it consistently outperformed other activation functions in terms of anomaly correlation coefficient (ACC) and root mean square error (RMSE) metrics. To reduce the problem of vanishing or exploding gradients during backpropagation, a problem often faced in deep neural networks, the ConvCastNet U-Net architecture is equipped with encoder-decoder skip connections (see Figure \ref{fig: ConcCastNet}). Additionally, each block features shared source skip connections, where the block input is concatenated with each subblock output. Input tensors are padded using the so-called \textit{spherical padding} (Figure \ref{fig: SphericalPadding}), which is designed to ensure that convolutional kernels cover a compact patch of Earth's surface, even if the kernel is positioned on the spatial boundary of the input tensor. The adverse effects of changes in the input distributions during training, i.e., covariate shift, are mitigated by batch normalisation. Our model has a total of 88 million weights.

We train the model using Adam optimiser \citep{Adam} with batch size 5 and starting learning rate $0.001$. We have observed that the small batch size results in faster convergence and smaller final loss. In addition, we utilise the ReduceLROnPlateau scheduler to adapt the learning rate during training. Specifically, the learning rate is reduced by a factor of $100$ following six epochs without observable improvement in the loss function. The loss function is $0.1 \cdot \mathrm{MSE}(y_{0,1,...,n},\hat{y}_{0,1,...,n})^\frac{1}{256}$, with $\mathrm{MSE}$ denoting the mean squared error, $y_{0,1,...,n}$ denoting $n$ consecutive target and $\hat{y}_{0,1,...,n}$ $n$ consecutive model prediction tensors. We change $n$ from 1 to 4 during training, thus gradually increasing the number of autoregressive steps used for error minimisation. The MSE exponent was treated as a hyperparameter, and the value $1/256$ was empirically determined to produce the best results on the validation dataset.
\begin{table*}[t]%
\centering
\caption{Chronologically listed training parameters showing the number of autoregressive steps used in loss minimisation $n$, number of epochs and time intervals for training and validation subsets. \label{tab: training}}%
\begin{tabular*}{\textwidth}{@{\extracolsep\fill}cccc@{\extracolsep\fill}}%
\toprule
\textbf{$n$} & \textbf{epochs} & \textbf{training} & \textbf{validation} \\
\midrule
1 & 20 & 1970-1984 & 2015-2019  \\
1 & 20 & 1985-1999 & 2015-2019  \\
1 & 20 & 2000-2014 & 2015-2019  \\
2 & 18 & 1980-1999 & 2015-2019  \\
2 & 20 & 2000-2014 & 2015-2019  \\
4 & 18 & 1985-1999 & 2015-2019  \\
4 & 20 & 2002-2017 & 2018-2019  \\
\bottomrule
\end{tabular*}
\end{table*}

\section{Long-range spatial error distribution}
\label{App: LongRangeSpatialErrorDistribution}
\vspace{12pt}

\begin{figure*}
\centerline{\includegraphics[width=1. \textwidth]{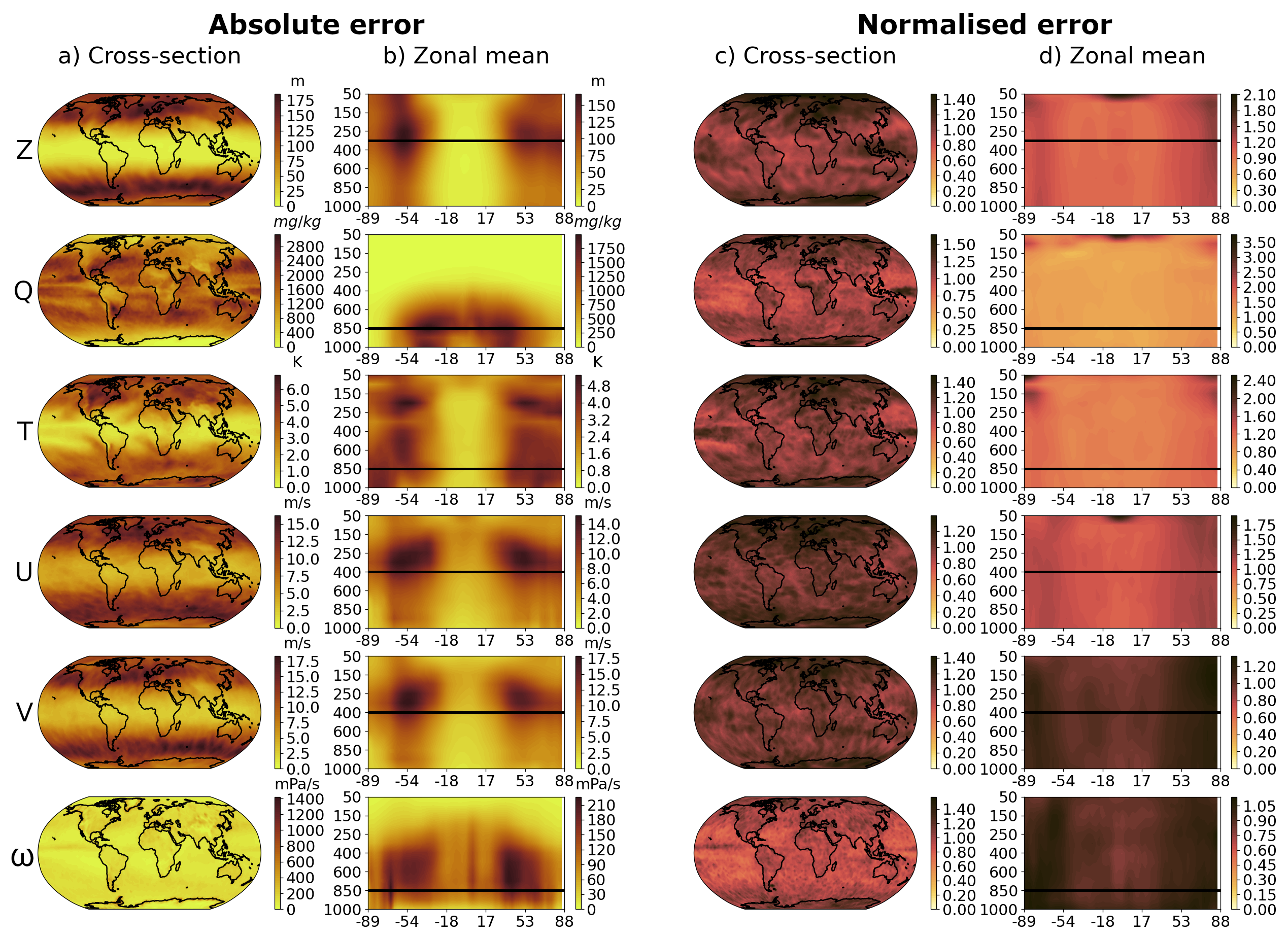}}
\caption{Same as Figure~\ref{spatial_errors_time_step_4}, but for 10-day lead time. \label{spatial_errors_time_step_20}}
\end{figure*}

Figure~\ref{spatial_errors_time_step_20} presents the spatial distribution of forecast errors at a 10-day lead time. Compared to the 2-day distribution (Fig.~\ref{spatial_errors_time_step_4}), the primary difference lies in the normalised error pattern, which evolves from being predominantly localised in the tropics to encompassing substantial errors in the extratropics. This shift is likely associated with the slower development of errors linked to the longer timescales of Rossby wave propagation.

\section{Non-standardised sensitivity}
\label{App: Non-standardised sensitivity}
\vspace{12pt}

Our sensitivity study is based on standardised errors and standardised input fields. When using non-standardised errors and inputs, Equation~\ref{eq: error} reads:
\begin{equation*}
 E_X = \sum_{v\in\mathcal{V}} \sum_{\mathbf{r} \in \mathcal{D}} \left| \hspace{1mm} \mathbf{X}_v^t(\mathbf{r}, t) - \mathbf{X}_v^f(\mathbf{r}, t) \right| \, ,
\end{equation*}
where $E_X$ denotes the non-standardised error. Using Eq.~\ref{eq: standardisation}, we can express the absolute error in terms of standardised fields: $ \left| \mathbf{X}_v^t - \mathbf{X}_v^f \right|= (\mathrm{std}(\mathbf{X}_v^c) + \mathbf{\varepsilon})\left| \hspace{1mm} \mathbf{S}_v^t - \mathbf{S}_v^f \right| $. 
Using Eq.~\ref{eq: standardisation} and denoting $\mathbf{\sigma}_v^c = (\mathrm{std}(\mathbf{X}_v^c) + \mathbf{\varepsilon})$, we show that the derivative of the standardised error with respect to standardised inputs is equivalent to the derivative of the non-standardised error with respect to non-standardised inputs:
\begin{equation*}
     \frac{\partial E_X} {\partial \mathbf{X}} = \left(\frac{\partial E_X} {\partial \mathbf{S}}\right) \frac{\partial \mathbf{S}}{\partial \mathbf{X}} = \left(\frac{\partial E_S} {\partial \mathbf{S}} \mathbf{\sigma}_v^c \right) \frac{1}{\mathbf{\sigma}_v^c} = \frac{\partial E_S} {\partial \mathbf{S}} ,
\end{equation*}
where $E_S$ and $E_X$ refer to the standardised and non-standardised errors, respectively. Our sensitivity analysis thus holds for destandardised fields as well.

\end{document}